\documentclass[10pt,pra,twocolumn,superscriptaddress,showkeys]{revtex4-1}
\usepackage{amsmath,graphicx,amsfonts,bm,amssymb}

\begin{document}

\title{Simulation of the Majorana equation in circuit QED}

\author{Sheng Liu}
\affiliation{Laboratory of Quantum Engineering and Quantum Materials,  South China
Normal University, Guangzhou 510006, China}
\affiliation{School of Physics and Telecommunication Engineering, South China
Normal University, Guangzhou 510006, China}

\author{Chuan-Jia Shan}
\affiliation{College of Physics and Electronic Science, Hubei Normal University, Huangshi 435002,
China}

\affiliation{Laboratory of Quantum Engineering and Quantum Materials,  South China
Normal University, Guangzhou 510006, China}

\author{Zhi-Ming Zhang}
\affiliation{Laboratory of Photonic Information Technology, and SIPSE, South China Normal University, Guangzhou 510006, China}

\affiliation{Laboratory of Quantum Engineering and Quantum Materials,  South China
Normal University, Guangzhou 510006, China}

\author{Zheng-Yuan Xue}\email{zyxue@scnu.edu.cn}
\affiliation{Laboratory of Quantum Engineering and Quantum Materials,  South China
Normal University, Guangzhou 510006, China}
\affiliation{School of Physics and Telecommunication Engineering, South China
Normal University, Guangzhou 510006, China}

\date{\today}

\begin{abstract}

We propose a scheme to simulate the 1D Majorana equation with two Cooper pair boxes coupled to a 1D superconducting
transmission line resonator, where strong coupling  limit can be achieved. With  proper choice of systematic parameters,
we are able to engineer different kind of interactions, which are indispensable for simulating the Majorana
equation in an enlarged real Hilbert space. Measurement of a conserved observable, i.e., the pseudo-helicity, via transmission spectrum of  the  cavity field can verify the simulated Majorana wave function. The measurement is experimentally resolvable according to our estimation  based on
conservative experimental parameters.
\end{abstract}

\pacs{03.67.Lx, 42.50.Dv, 74.78.Na}

\keywords{Quantum simulation, Majorana equation,   Circuit QED}

\maketitle


Quantum simulators \cite{qs1,qs2,cirac} can be used to  study quantum systems that are
beyond the reach of classical computers. Meanwhile, they are expected to be more robust against various imperfections than quantum computers \cite{cirac}. One of the successful cases is the  quantum simulation of the Dirac equation, which combines  quantum mechanics and special relativity \cite{Thaller}. In certain regime, electrons in graphene may behavior as Dirac fermions, which have recently raised strong interest in condensed matter physics \cite{kane}. Meanwhile, it is proposed that ultracold atoms in an optical lattice can be used to simulate such relativistic Dirac fermions \cite{zhu}. Comparing with graphene, the atomic simulation may offer more controllability. Alternatively, simulation of the Dirac equation is also proposed in cold atoms \cite{gauge} with light-induced gauge potential \cite{gauge1,zhugauge}. Meanwhile, quantum simulation of the Dirac equation with trapped ions has also been proposed \cite{L.Lamata,L.Lamata2} and experimentally verified \cite{R.Gerritsma}.

One of the greatest success of the applications of the Dirac equation is that it predicts the existence of antiparticle for electron. In viewing of the success, Majorana inquired whether it is  possible that a  particle to be its own antiparticle. As a result, he found an  equation that such particles  should satisfy, i.e., the Majorana equation (ME) \cite{Wilczek}. Recently, it is  proposed that ME can be simulated with trapped ions \cite{Casanova}, which is developed in an
enlarged Hilbert space: An 1D ME is transformed to a 3D Dirac equation with dimensional reduction, i.e., the momenta in $y$ and $z$ directions of which are zero. The simulation of the ME is not straightforwardly as it is non-Hermitian, i.e., one needs to  implement the  complex conjugation of the Majorana wave function. Therefore, it needs new toolbox to access this unphysical operation in the simulation.

Superconducting system is generally regarded as one of the most promising
candidates for physical implementation of qubit that can support
scalable quantum information processing \cite{qc1,qc2,qc3,qc4,qc5}.
Recently, quantum simulators using superconducting circuits have attracted much attention \cite{houck,qscqed}. With  mature microchip fabrication techniques, it can be used to simulate quantum systems in a  flexible way. Especially,
quantum simulation of the dynamical Casimir effect  \cite{r1,r2,r3,r4,r5}, topologically protected states \cite{s1,s2,xue1,xue2,xue3}, and single-photon transport \cite{s3,s4,s5,s6} are explored with superconducting circuits.

Here, we consider  simulation of the ME with two Cooper pair boxes
(CPB) coupled to microwave fields \cite{you,zhu1,zhu2,xue4}. Experimentally,
this can be achieved by CPB capacitively coupled to a 1D transmission line resonator (TLR) \cite{A.Wallraff}. Therefore, our scheme provides an interesting example of simulating physics associated with the relativistic ME in a  mesoscopic circuit. As shown in Ref. \cite{Casanova}, the simulation is based on complex-to-real map, which transforms a ME into a higher-dimensional Dirac equation. For 2D or 3D ME
simulation, one needs more qubits (more than 3) to simulate the
transformed Hamiltonian. Unfortunately, in trapped ions system,
individual addressing is difficult for large arrays ($N>3$) \cite{ion2}.
Comparing with the trapped ion simulation, a distinct feature of the present proposal is that the combination of individual addressing with a many CPB setup is feasible.


A ME  reads
\begin{equation}
\label{U}
i\hbar \gamma^\mu\partial_\mu\psi=mc\psi_c,
\end{equation}
where  $c$ is the  speed of light, $m$ is the mass, $\gamma^\mu$
are the Dirac matrices with $\mu$=1, 2, 3, and 4,
and $\psi_c=i\gamma^2\mathcal{K}\psi$ is a charge conjugation field with
$\mathcal{K}$ being a complex conjugation operator. For 1D case,
$\psi_c=\mathcal{C}\sigma_z\psi^*$, where  $\mathcal{C}$ is a
unitary matrix satisfying $\mathcal{C}\gamma^T=-\gamma\mathcal{C}$
with $\gamma=i\sigma_x\sigma_z$.  We can choose
$\mathcal{C}=i\sigma_y$ in a suitable basis, i.e.,
$\psi_c=i\sigma_y\sigma_z\mathcal{K}\psi=i\sigma_y\sigma_z\psi^*$. Then, the ME in 1D reads
\begin{equation}
\label{A}
i\hbar\partial_t\psi=(c\sigma_xp_x-imc^2\sigma_y\mathcal{K})\psi,
\end{equation}
where $\psi$ is a two-component complex spinor, and $p_x=-i\hbar\partial_x$ is the momentum operator in $x$ direction.

To make the unphysical complex conjugation $\psi\rightarrow\mathcal{K}\psi=\psi^*$ to be implementable, one can map the two-component complex
spinor into a four-component real spinor  \cite{Casanova}, that is,
\begin{equation}
 \psi=
 \left(\begin{array}{c}
 \psi_1\\
 \psi_2
 \end{array}\right)\in\mathbb C_2
\rightarrow\Psi=\frac{1}{2}
 \left(\begin{array}{c}
 \psi+\psi^*\\
 i(\psi^*-\psi)
 \end{array}\right)\in\mathbb{R}_{4}.
\end{equation}
After this map, one can unify all the antiunitary or unitary
operators and complex field $\psi$ in an enlarged real Hilbert space.
Then,  Eq. (\ref{A}) reads
\begin{equation}
\label{B}
i\hbar\partial_t\Psi=[c(\mathbb
E\otimes\sigma_x)p_x-mc^2\sigma_x\otimes\sigma_y]\Psi,
\end{equation}
where $\Psi=  \left(
 \psi^r_1 \quad
 \psi^r_2 \quad
 \psi^i_1 \quad
 \psi^i_2
 \right)^T$ and it now becomes an implementable Hamiltonian equation.
In Eq. (\ref{B}), the four components of the spinor are nonlinearly coupled and cannot be separated, so it is a  $1+1$ Dimension ME with an irreducible 4 Dimensional Hilbert space notation. In the following, we will  simulate the dynamics of the ME in 1+1 Dimension by engineering this Hamiltonian. Meanwhile, it is well known that the ME in 3D conserves the observable of helicity. In 1D, helicity reduces to the so-called pseudo-helicity $\Sigma=\sigma_xp_x$, and it is still conserved. However, it is not conserved in 1D Dirac equation 
as $[\Sigma, H_D]\neq0$,  where $H_D=c\sigma_xp_x+mc^2\sigma_z$ is the 1D Dirac Hamiltonian. Therefore, measurement of the pseudo-helicity can demonstrate the Majorana wave function and thus verify our dynamical simulation. Mapping into the real Hilbert space, the pseudo-helicity $\Sigma$ reads
\begin{equation}
\tilde{\Sigma}=M^\dagger\sigma_xp_xM=(\mathbb{E}\otimes\sigma_x-\sigma_y\otimes\sigma_x),
p_x
\end{equation}
where $\mathbb{E}$ is identity matrix.



We now turn to our circuit QED simulation. The qubit considered here
is the superconducting CPB consisting of superconducting island where  two Josephson  junctions  with capacitance $C_J$ and Josephson energy $E_j$ are configured into a
loop geometry, which is pierced by an external applied  magnetic flux
$\Phi$. When the Josephson energy is much smaller than the
charging energy $E_c=e^2/2C_{\Sigma}$ $(C_{\Sigma}=C_g+2C_J)$ and
restricting the induced charge $N_g=C_gV_g^{dc}/(2e)$ within the range of
$N_g\in[0,1]$, only a pair of adjacent charge states on the inland
are relevant. Then, the CPB  reduces to a simple two-level system described by \cite{qc1}
\begin{equation}
\label{C}
H_a=-\frac{E_{el}}{2}\bar{\sigma}_z-\frac{E_J}{2}\bar{\sigma}_x,
\end{equation}
where $\bar{\sigma}_x$, $\bar{\sigma}_z$ are the pauli matrices in
the charge basis of $\{|0\rangle$, $|1\rangle\}$, $E_{el}=4E_c(1-2N_g)$ is
the electrostatic energy, and $E_J=2E_j\cos(\pi \Phi/\Phi_0)$ is the
effective Josephson energy with $\Phi_0$ being the flux quanta. From Eq. (\ref{C}), one can see that it is possible and convenient to control the qubit
by the applied gate voltage $V_g^{dc}$ and the pierced flux $\Phi$
\cite{qc1}. Therefore, the qubit splitting energy can be
tunable by the external magnetic flux even with fix gate voltage,
e.g., at the degeneracy point $N_g=1/2$.

In  circuit QED, the CPB is capacitively coupled to the center
conductor via a capacitance $C_g$, at the cavity mode's antinode with  a maximum voltage. Meanwhile, besides the dc control voltage $V_g^{dc}$, the gate voltage on the CPB also includes an ac part from the oscillating cavity mode. Taking both parts into
consideration,   Eq. (\ref{C}) reads
\begin{eqnarray}
\label{D}
H_{int}&=&-2E_c\left(1-2N_g^{dc}\right)\bar{\sigma}_z-\frac{E_J}{2}\bar{\sigma}_x\notag\\
&&+\hbar w_ra^\dagger a-\hbar g\left(a+a^\dagger\right)
\left(1-2N_g^{dc}-\bar{\sigma}_z\right),
\end{eqnarray}
where $a^\dagger$ and $a$ are the  creation and annihilation operators
of the cavity mode, $w_{r}$ is its frequency, and
$g$ is the tunable coupling strength with $g/2\pi \in[5.8, 100]$ MHz
\cite{A.Blais}. Denoting $\{|\downarrow\rangle,
|\uparrow\rangle\}$ as the ground and  excited states of
the first two terms of Hamiltonian (\ref{D}), respectively. In this
new basis, at the degeneracy point,  within  the rotating-wave
approximation, Hamiltonian (\ref{D})  reduces to the
Jaynes-Cummings form \cite{A.Blais}
\begin{equation}
\label{G}
H_{JC}=\hbar w_ra^\dagger a+\frac{\hbar\omega}{2}\sigma_z+\hbar
g(a^\dagger\sigma_-+a\sigma_+)
\end{equation}
where  $\omega=E_J/\hbar$ and $\sigma_z$ is pauli matrices in the new basis.


In addition, a driven microwave field of frequency $w_d$ can also
be capacitively coupled to the resonator, which can be
in the form of
\begin{equation}
\label{H}
H_d(t)=\hbar\varepsilon(t)\left[a^\dagger
e^{-i(w_dt-\phi)}+ae^{i(w_dt-\phi)}\right],
\end{equation}
where $\varepsilon(t)$ and  $\phi$ are the amplitude and initial
phase of the driven microwave field,  respectively. All of the local
operations on the qubit are rely on  $w_d$, $\varepsilon(t)$ and phase
$\phi$, which have been experimentally achieved \cite{A.Wallraff}. When the driven
amplitude is  large, comparing with the vacuum fluctuation of the resonator, the
microwave field can be treated  as a classical field. Make a unitary
transformation $U=\exp(\alpha a^\dagger-\alpha^*a)$ on the total
Hamiltonian  consists of $(\ref{G})$ and $(\ref{H})$ leads to \cite{xue}
\begin{equation}
\label{I}
\begin{array}{l}\displaystyle
 H_{DJC1}=\hbar w_r a^\dagger a+\frac{\hbar\omega}{2}\sigma_z+\hbar
 g\left[(a^\dagger+\alpha^*)\sigma_-+H.c.\right].
\end{array}
\end{equation}
Assuming that the driven amplitude is independent of time, one obtains
$-\alpha\Delta=\varepsilon
\exp[-i(w_dt-\phi)]$ with  $\Delta=\omega_r-\omega_d$. In the
rotating frame at frequency $w_d$, we can rewrite $(\ref{I})$ as
\begin{eqnarray}
\label{K}
H_{DJC2}&=&\hbar\Delta a^\dagger a+\hbar g(a^\dagger
\sigma_-+a\sigma_+) +\frac{\hbar\delta^\prime}{2}\sigma_z \nonumber\\
&&+\frac{\hbar\Omega_d}{2}(\sigma_-e^{-i\phi}+\sigma_+e^{i\phi}),
\end{eqnarray}
 where
$\Omega_d=2g\varepsilon/\Delta$ and $\delta^\prime=\omega-\omega_d$.

Working in the eigenbasis of the  last two terms of the above
Hamiltonian, the Hamiltonian $(\ref{K})$ becomes
\begin{eqnarray}
\label{l}
H_{DJC3}&=&\hbar\Delta a^\dagger a+\frac{\hbar\Omega}{2}\sigma_z
+\frac{\hbar g}{2}[ae^{-i\phi}(\cos(\theta^\prime)\sigma_z \notag\\
&&-\sin(\theta^\prime)\sigma_x +\sigma_+-\sigma_-)+H.c.],
\end{eqnarray}
where $\theta^\prime=\arctan(\delta^\prime /\Omega_d)$ and
$\Omega=\sqrt{\Omega_d^2+{\delta^\prime}^2}$.
In the interaction picture, the interaction Hamiltonian of $(\ref{l})$ reads \cite{xue,s}
\begin{equation} \label{Tmn}
 H_I= \frac{\hbar g}{2}[ae^{-i\Delta
t}e^{-i\phi}(\sigma_z +\sigma_+e^{i\Omega
t}-\sigma_-e^{-i\Omega
t})+\text{H.c}.],
\end{equation}
which is key to achieve our effective Hamiltonian for simulating the ME
and $\delta^\prime=2n\pi\Omega_d$   with $n$
being an integer. From this driven Hamiltonian, one can engineer different type
of interactions. For $\Delta=\Omega$,  one obtains
\begin{equation}
 H_I^\prime=\frac{\hbar g}{2}a  e^{-i\phi}(\sigma_+ +\sigma_ze^{-i\Omega
t}-\sigma_-e^{-2i\Omega t})+\text{H.c}..
\end{equation}
In rotating-wave approximation, one can neglect the fast oscillating terms, then
the above Hamiltonian reduces to
\begin{equation}
\label{M}
H_1=\frac{\hbar g}{2}(a\sigma_+e^{-i\phi}+a^\dagger\sigma_-e^{i\phi})
\end{equation}
Similarly, for $\Delta=-\Omega$, one obtains
\begin{equation}
\label{N}
H_2=-\frac{\hbar g}{2}(a\sigma_-e^{-i\phi}+a^\dagger\sigma_+e^{i\phi})
\end{equation}
Meanwhile, for the strong driven case $\Omega\gg \Delta$, one obtains
\begin{equation}
\label{P}
H_3=\frac{\hbar g}{2}(ae^{-i(\Delta t+\phi)}+a^\dagger e^{i(\Delta
t+\phi)})\sigma_z.
\end{equation}

\begin{figure}
\begin{center} \includegraphics[width=8cm]{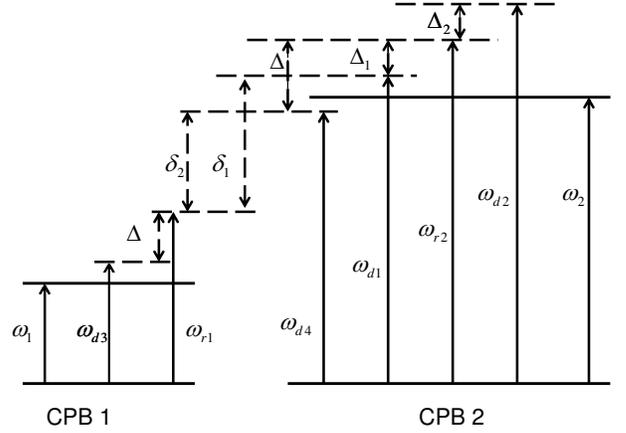} \end{center}
\caption{Level structure of our scheme. $\omega_{1}$, $\omega_{2}$
are level separation of the two CPBs, the frequency of the four
microwave pulses are $\omega_{d1}, \omega_{d2}$, $\omega_{d3}$ and
$\omega_{d4}$, and $\omega_{r1}$ and $\omega_{r2}$ are the frequency
of the two cavity modes.}
\end{figure}


To simulate the 1D ME in circuit QED,  we use two CPBs with level
separation $\omega_{1}$ and $\omega_{2}$, four microwave pulses of
frequency $\omega_{d1}, \omega_{d2}$, $\omega_{d3}$ and
$\omega_{d4}$ with initial phase $\phi_1$, $\phi_2$, $\phi_3$ and
$\phi_4$, and two cavity modes of frequency $\omega_{r1}$ and
$\omega_{r2}$. The level structure  and the frequency of the driven
fields are shown in Fig. 1. As explicitly  shown in Eq. (\ref{B}),
there are two components: the first one is the kinetic term of CPB 2
$cp_x(\mathbb{E}\otimes\sigma_x)$ and the second one is the exchange
coupling of the two CPBs. As the kinetic term is the combination of
Hamiltonian (\ref{M}) and (\ref{N}),  it can be generated by two
detuned driving microwave fields of frequencies $\omega_{d1}$ and
$\omega_{d2}$ with detuning
$\Delta_1=\omega_{r2}-\omega_{d1}=\Omega_1$ and
$\Delta_2=\omega_{r2}-\omega_{d2}=-\Omega_2$, and the initial phases
$\phi_1=\pi/2$ and $\phi_2=-\pi/2$. In typical experiments \cite{A.Wallraff,aw2}, we can choose
$\omega_{r2}=10$ GHz, $\omega_{d1}=9.9$ GHz, $\omega_{d2}=10.1$ GHz,
and $\Omega_1=\Omega_2=100$ MHz. Then, we get the combined  kinetic term
Hamiltonian $i\hbar g(a^\dagger-a)(\mathbb{E}\otimes\sigma_x)/2$ for CPB 2 with level separation of $\omega_2=9.7$ GHz. Meanwhile, we on purposely chosen the two CPBs with much different energy splitting with $\omega_1=4.4$ GHz, and thus
$\omega_{d1}$  and  $\omega_{d2}$  cannot generate similar kinetic
term on CPB 1 as $\delta_1=\omega_{d1}-\omega_{r1}=4.9$ GHz  is much larger than
$ \{\Delta_1, |\Delta_2|\}=100$ MHz, where we have chosen $\omega_{r1}=5$ GHz.
The second term is generated by virtual excitation of the TLR, where we
need the same detuning between the driving microwave fields and the two
CPBs. As we choose the different CPB splitting energy, we also need
two additional strong driving microwave field: $\{\Omega_3,
\Omega_4\}\gg\Delta=\omega_{r1}-\omega_{d3}=\omega_{r2}-\omega_{d4}=500$ MHz,
where we have chosen the microwave pulse with frequencies $\omega_{d3}=4.5$ GHz and $\omega_{d4}=9.5$ GHz. Using the effective Hamiltonian (\ref{P}) with initial phase $\phi_3=\phi_4=0$, we obtain $\frac{\hbar
g}{2}(\sigma_z\otimes\mathbb{E}+\mathbb{E}\otimes\sigma_z)(a^\dagger
e^{i\Delta t}+ae^{-i\Delta t})$.  The cross talk between the two coupling
channels on CPB2, kinetic and exchange coupling, can be eliminated via
rotating-wave approximation because of the chosen parameters $\Delta \gg \Delta_1$. Meanwhile, the driven on CPB 1 by $\omega_{d4}$ is neglecting small while
$\delta_2=\omega_{d4}-\omega_{r1}\gg \Delta$. Then, in the
interaction picture, the complete Hamiltonian reads
\begin{eqnarray}
\label{Q}
H&=&\frac{\hbar g}{2}(\sigma_z\otimes\mathbb{E}+\mathbb{E}\otimes\sigma_z)(a^\dagger
e^{i\Delta t}+ae^{-i\Delta t}) \nonumber\\
&&+\frac{\hbar g}{2}i(a^\dagger-a)(\mathbb{E}\otimes\sigma_x).
\end{eqnarray}
Rotating on the CPB1 and CPB2,  Hamiltonian (\ref{Q}) can be rotated to
\begin{eqnarray}
\label{R}
H&=&\frac{\hbar
g}{2}(\sigma_x\otimes\mathbb{E}+\mathbb{E}\otimes\sigma_y)(a^\dagger
e^{i\Delta t}+ae^{-i\Delta t})   \nonumber\\
&&+\frac{\hbar g}{2}i(a^\dagger-a)(\mathbb{E}\otimes\sigma_x),
\end{eqnarray}
the effective Hamiltonian of which recover the 1D ME Hamiltonian in Eq. (\ref{B})
with the identification of
\begin{equation}
p_x=i\sqrt{\frac{m^\prime \hbar\omega_{r2}}{2}}(a^\dagger-a), \quad
c=g\sqrt{\frac{\hbar}{2m^\prime w_{r2}}}, \quad mc^2=\frac{\hbar
g^2}{2\Delta},
\end{equation}
where $m^\prime$ is the  mass of inductance.



As mentioned above, to verify our simulation, we need to measure the
pseudo-helicity, which is conserved in the 1D ME. We now move to measure the pseudo-helicity in circuit QED. We first deal with the kinetic term, which can be measured as following: (1) a state-dependent operator $U_2=\exp(-ik(\mathbb{E}\otimes\sigma_y)\otimes p_x/2)$ acting on
CPB2, which can be  generated by two microwave pulses with both
initial phases being $\pi$; (2) a $\sigma_z$ measurement of CPB2,
which can be measured by microwave irradiation of the cavity and
then probing the transition frequency to determine the qubit state \cite{R.Gerritsma,A.Blais}. The above two steps
equal to the measurement of \cite{Casanova}
\begin{eqnarray}
F(k)&=&U_2^\dagger(\mathbb{E}\otimes\sigma_z)U_2 \nonumber\\
&=&\cos(kp_x)(\mathbb{E}\otimes\sigma_z)-\sin(kp_x)(\mathbb{E}\otimes\sigma_x),
\end{eqnarray}
where $k$ is determined by the probing time \cite{R.Gerritsma}. As
$\left.\frac{d}{dk}\langle F(k\rangle)\right|_{k=0}=-\langle(\mathbb
E\otimes\sigma_x)\otimes p_x\rangle $, the kinetic term of Eq.
$(10)$ can be measured by probing the initial slope of the
observable $F(k)$ \cite{R.Gerritsma}. Similarly, the second spin
exchange interaction term in Eq. (10), can be measured by  a unitary
transformation $U_1=\exp(-ik(\sigma_x\otimes\mathbb{E})\otimes
p_x/2)$ on CPB1, and measure the spin correlation $\sigma_z\otimes\sigma_x$. For the former operation, an additional $\pi/2$ pulse  is needed. Then, probing the initial slope as before, we obtain
 $\left. \frac{d\langle\sigma_z\otimes\sigma_x\rangle}{d
 k}\right|_{k=0}=2\langle(\sigma_y\otimes\sigma_x)\otimes  p_x\rangle$.
Combining both the kinetic and spin exchange terms, we  achieve our
final goal of measuring the pseudo-helicity.


Although the absolute value of the pseudo-helicity is small, it should
not hamper our measurement since what we need  to measure is $\sigma_z$. In circuit QED, high-fidelity quantum non-demolition measurements of $\sigma_z$ is now being an experimental routine \cite{A.Blais}.  In the  following, we highlight the measurement
process in our scheme with input-output formulism. In our model, as shown in Fig. 2, if the cavity contains two qubits, the intracavity field come from $b_{in}$ will acquire some non-trivial dynamics which then compel the external field $a_{out}$ to have a time dependent difference comparing with the free field dynamics.  As a result, there is a qubit-state-dependent phase shift between the input and  output fields \cite{D.F.}.

\begin{figure}\begin{center}
\includegraphics[width=7cm,]{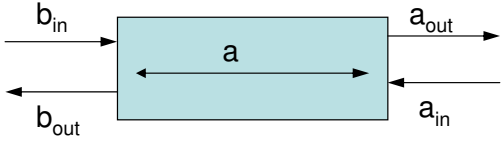} \end{center}
\caption{A schematic representation of the cavity field, input field,
and output field  for a two-side leaky cavity.}
\end{figure}

As we all know that there is  loss mechanism in all physical processes,
particularly for measurements, which strongly influence the precision of the quantum
non-demolition  measurement. Therefore, dispersive regime should be considered in our measurements. Insight into the dispersive regime between  the CPB $1$ and the intracavity can be obtained by a unitary transformation
\begin{equation}
\label{V}
U_1=\exp\left[\frac{g}{\Delta^\prime}(a\sigma_+-a^\dagger\sigma_-)\right],
\end{equation}
where a large detuning $\Delta^\prime=\omega_{r1}-\omega_1\gg g$ is assumed. Applying transformation (\ref{V}) on the  Hamiltonian (\ref{G})  and (\ref{H}) to second order in $g$ (neglecting the damping for the moment), for CPB 1, we can get
\begin{eqnarray}
H_{sys1}&=& U_1(H_{JC}+H_d)U_1^\dagger \nonumber\\
&\approx &\hbar(\omega_{r1}-\omega_{d3}) a^\dagger a
+\frac{\hbar}{2}(\omega_1-\omega_{d3})\sigma_z+\frac{\hbar
g^2}{\Delta^\prime}a^\dagger a\sigma_z\nonumber\\
&&+\frac{\hbar g^2}{2\Delta^\prime}\sigma_z+\frac{\hbar
g\varepsilon}{\Delta^\prime}\sigma_x+\hbar\varepsilon(a^\dagger+a)
\end{eqnarray}
in a rotating frame at the driven frequency $\omega_{d3}$.

Consider the resonator as a two-side leaky cavity with equal
rates, as shown in Fig. 2, the relationship between the input and
output modes can be written as $a_{out}(t)=\sqrt{\kappa}a(t)-a_{in}(t)$,
where $a_{out}$ and $a_{in}$  are the output and input modes at
the output port, respectively; $a$ is the  intracavity mode, and cavity lifetime
$1/\kappa$ is about $160$ ns in  a typical circuit QED system \cite{A.Wallraff}.  The quantum voltage is related to the current carried by the TLR  by
$I(x,t)=\sqrt{\frac{c}{l}}V(x,t)$. For the moment, it is more convenient to have stationary rather than traveling quantum voltage.   For this mode, we obtain the following equation of motion for the CPB $1$
\begin{eqnarray}
\frac{da(t)}{dt}&=&-\frac{i}{\hbar}[a(t),H_{sys1}]-\kappa
a(t)+\sqrt{\kappa}a_{in}(t)+\sqrt{\kappa}b_{in}(t)\nonumber\\
&=&-i[(\omega_{r1}-\omega_{d3})a+\varepsilon+\chi_1 a\sigma_z] \nonumber\\
&-&\kappa a(t)   +\sqrt{\kappa}a_{in}(t)+\sqrt{\kappa}b_{in}(t),
\end{eqnarray}
where $\chi_1=g^2/\Delta^\prime$ and $b_{in}$ is the input mode at
the input port of the resonator. After the Fourier transformation, we can obtain a spectra function between the input and output modes as
\begin{eqnarray}
a_{out}(\omega)&=&\frac{\kappa\left[a_{in}(\omega)+b_{in}(\omega)\right]
-i2\pi\sqrt{\kappa}\varepsilon\delta(\omega)}
{\kappa+i(\chi_1\sigma_z+\omega_{r1}-\omega_{d3}-\omega)}-a_{in}(w)\nonumber\\
\end{eqnarray}
where $\delta(\omega)$ is the Dirac function. Obviously, the
$\sigma_z$ in the denominator is a formal indication that the output
spectra should depend on the qubit state ($\sigma_z=\pm1$). In the case of
$\omega=\omega_{r1}-\omega_{d3}$, we can get \cite{Mohan}
\begin{equation}
a_{out}^\mp(w_{r1}-\omega_{d3})=\frac{\kappa b_{in}(\omega_{r1}-\omega_{d3})}{\kappa\pm
i\chi_1}\mp\frac{ i\chi_1}{\kappa\pm\chi_1}a_{in}(\omega_{r1}-\omega_{d3}).
\end{equation}
Assuming the two input modes are independent and  $a_{in}(t)$ is in the vacuum state because of  very small reflection and backscattering of the resonator. Neglecting the
two-photon process, $a_{in}(w)$ has no contribution to the normally ordered moment $a_{out}(w)$, and thus,
\begin{eqnarray}
\langle
a^\mp_{out}(\omega_{r1}-\omega_{d3})\rangle_{\mathcal{N}1}&=&\left\langle\frac{\kappa
b_{in}(\omega_{r1}-\omega_{d3})}{\kappa\pm
i\chi_1}\right\rangle_{\mathcal{N}1}     \nonumber\\
&=&\langle  e^{i\theta_1^{\pm}}\tilde{b}_{in}(\omega_{r1}-\omega_{d3})\rangle_{\mathcal{N}1},
\end{eqnarray}
where  $\langle\cdot\rangle_{\mathcal{N}}$ signify a normally
ordered moment, $\theta_1^{\pm}$ is the phase shift dependent on the qubit state between the input field $b_{in}$ and the output field $a_{out}$ [$\tan(\theta _1^{\pm})=\mp\frac{g^2}{\Delta^\prime\kappa}$], and $\tilde{b}_{in}$ is a scaled quantity of $b_{in}$. When the cavity resonance frequency
and the driven frequency are confirmed, the phase shift depends on the CPB1 qubit  state is about $\pm \frac{134\pi}{360}$ and the corresponding frequency interval is $29.6$ MHz for our scheme, which is readily resolvable experimentally \cite{A.Wallraff}.

Similarly, analogical transformation can be applied on CPB2,    we  can obtain
\begin{eqnarray}
\langle
a^\mp_{out}(\omega^\prime)\rangle_{\mathcal{N}2}=\left\langle\frac{\kappa
b_{in}(\omega^\prime)}{\kappa\pm
i\chi_2}\right\rangle_{\mathcal{N}2}
=\langle e^{i\theta_2}\tilde{b}_{in}(\omega^\prime)\rangle_{\mathcal N2},
\end{eqnarray}
where $\omega^\prime=\omega_{d1}+\omega_{d2}+\omega_{d4}-\omega_{r2}$,
$\tan(\theta_2^\pm)=\frac{\mp g^2}{\Delta^{\prime\prime}\kappa}$ and
$\chi_2=g^2/\Delta^{\prime\prime}$ with
$\Delta^{\prime\prime}=\omega_{r2}-\omega_2\gg g$. For the CPB2,
$\theta_2=\pm\frac{268\pi}{360}$ and the frequency interval is $59.2$ MHz.

In summary, we propose to simulate the 1D ME with  two CPBs
capacitively coupled to a TLR in circuit QED system
where we are able to engineer different kinds of
interaction which constructing the wanted Hamiltonian for the ME. The conserved observable pseudo-helicity is measured  via the input-output process, estimation based on conservative parameters shows that it is experimentally resolvable.

\bigskip

This work was supported by NFRPC (No. 2013CB921804, and No. 2011CB922104), NSFC (No. 60978009, and No.91121023), the PCSIRT (No. IRT1243), and the Zhongshan municipal scientific project (No. 20123A326).


\begin{thebibliography}{99}

\bibitem{qs1} Buluta I.,  Nori, F.:
Quantum simulators. Science \textbf{326}, 108-111 (2009)

\bibitem{qs2} Georgescu, I.M., Ashhab, S.,  Nori, F.:
Quantum simulation. Rev. Mod. Phys. \textbf{86}, 153-185 (2014)

\bibitem{cirac} Cirac, J. I.,  Zoller,  P.:
Goals and opportunities in quantum simulation.
Nat. Phys. \textbf{8}, 264-266 (2012)

\bibitem{Thaller} Thaller,  B.: The Dirac equation. Springer,
Berlin (1992)


\bibitem{kane} Kane, C.L.,  Mele, E.J.:
Quantum spin Hall effect in graphene.
Phys. Rev. Lett. \textbf{95}, 226801 (2005)


\bibitem{zhu} Zhu, S.-L., Wang, B.,   Duan, L.-M.:
Simulation and detection of Dirac fermions with cold atoms in an optical lattice.
Phys. Rev. Lett. \textbf{98}, 260402 (2007)


\bibitem{gauge}  Juzeli\={u}nas, G.,  Ruseckas,  J.,  Lindberg, M., Santos, L.,  \"{O}hberg, P.:
Quasirelativistic behavior of cold atoms in light fields.
Phys. Rev. A \textbf{77}, 011802 (2008)


\bibitem{gauge1} Ruseckas, J., Juzeli\={u}nas,  G., \"{O}hberg, P.,  Fleischhauer, M.:
Non-Abelian gauge potentials for ultracold atoms with degenerate dark states.
Phys. Rev. Lett. \textbf{95}, 010404 (2005)

\bibitem{zhugauge} Zhu,  S.-L., Fu, H., Wu, C.-J., Zhang, S.-C., Duan,  L.-M.:
Spin Hall effects for cold atoms in a light-induced gauge potential.
Phys. Rev. Lett. \textbf{97}, 240401 (2006)

\bibitem{L.Lamata} Lamata, L., Le\'{o}n, J., Sch\"{a}tz, T.,  Solano,  E.:
Dirac equation and quantum relativistic effects in a single trapped ion.
Phys. Rev. Lett. \textbf{98}, 253005 (2007)

\bibitem{L.Lamata2} Casanova, J., Garc\'{i}a-Ripoll, J.J., Gerritsma,  R., Roos, C. F.,  Solano,  E.:
Klein tunneling and Dirac potentials in trapped ions.
Phys. Rev. A \textbf{82}, 020101 (2010)

\bibitem{R.Gerritsma} Gerritsma, R., Kirchmair, G., Z\"{a}hringer, F., Solano, E., Blatt, R. ,   Roos,  C. F.:
Quantum simulation of the Dirac equation.
Nature (London) \textbf{463},  68-71 (2010)

\bibitem{Wilczek} F. Wilczek,: Majorana returns.
Nat. Phys. \textbf{5},  614-618 (2009)


\bibitem{Casanova} Casanova, J., Sab\'{\i}n, C., Le\'{o}n, J., Egusquiza, I.L., Gerritsma, R., Roos, C.F., Garc\'{i}a-Ripoll, J.J., Solano, E.:
Quantum simulation of the Majorana equation and unphysical operation.
Phys. Rev. X \textbf{1}, 021018 (2011)


\bibitem{qc1} Makhlin, Y.,  Sch\"{o}n, G.,  Shnirman, A.:
Quantum-state engineering with Josephson-jounction devices.
Rev. Mod. phys. \textbf{73}, 357-400 (2001)

\bibitem{qc2} You, J.Q., Nori, F.:
Superconducting circuits and quantum information.
Phys. Today \textbf{58}(11), 42-47 (2005)

\bibitem{qc3} You, J.Q., Nori, F.:
Atomic physics and quantum optics using superconducting circuits.
Nature (London) \textbf{474},
589-597 (2011)

\bibitem{qc4} Buluta, I., Ashhab, S., Nori, F.: Natural and artificial atoms for quantum computation. Rep. Prog. Phys.
\textbf{74}, 104401 (2011)

\bibitem{qc5} Xiang, Z.-L., Ashhab, S., You, J.Q., Nori, F.:
Hybrid quantum circuits: Superconducting circuits interacting with other quantum systems.
Rev. Mod. phys. \textbf{85}, 623-653 (2013)


\bibitem{houck} Houck,  A.A., T\"{a}reci, H.E.,  Koch,  J.:
On-chip quantum simulation with superconducting circuits.
Nat. Phys. \textbf{8}, 292-299 (2012).

\bibitem{qscqed} Shevchenko, S.N.,  Ashhab, S.,  Nori, F.:
Landau-Zener-St\"{u}ckelberg interferometry.
Phys. Rep. \textbf{492}, 1-30 (2010)


\bibitem{r1} Nation, P.D.,  Johansson, J.R.,  Blencowe, M.P.,  Nori, F.:
Stimulating uncertainty: Amplifying the quantum vacuum with superconducting circuits.
Rev. Mod. Phys. \textbf{84}, 1-24 (2012)

\bibitem{r2} Johansson, J.R., Johansson, G., Wilson, C.M.,  Nori, F.:
Dynamical Casimir effect in a superconducting coplanar waveguide. Phys. Rev. Lett. \textbf{103}, 147003 (2009)

\bibitem{r3} Johansson, J.R., Johansson, G., Wilson, C.M.,  Nori, F.:
Dynamical Casimir effect in superconducting microwave circuits. Phys. Rev. A \textbf{82}, 052509 (2010)

\bibitem{r4} Johansson, J.R., Johansson, G., Wilson, C.M., Delsing, P., Nori, F.:
Nonclassical microwave radiation from the dynamical Casimir effect. Phys. Rev. A \textbf{87}, 043804 (2013)

\bibitem{r5} Wilson, C.M., Johansson, G., Pourkabirian, A.,  Simoen, M., Johansson, J.R., Duty, T., Nori, F., Delsing,  P.:
Observation of the dynamical Casimir effect in a superconducting circuit.
Nature  (London)  \textbf{479}, 376-379 (2011)

\bibitem{s1}  You, J.Q., Shi, X.-F.,Hu, X.,   Nori, F.:
Quantum emulation of a spin system with topologically protected ground states using superconducting quantum circuits.
Phys. Rev. B \textbf{81}, 014505  (2010)

\bibitem{s2} You, J.Q., Wang, Z.D., Zhang, W., Nori, F.:
Manipulating and probing Majorana fermions using superconducting circuits.
arxiv:1108.3712 (2011)

\bibitem{xue1} Xue, Z.-Y., Wang, Z.D., Zhu, S.-L.: Physical implementation of topologically decoherence-protected
superconducting qubits. Phys. Rev. A \textbf{77}, 024301 (2008)

\bibitem{xue2}  Xue, Z.-Y., Zhu, S.-L.,  You, J.Q., Wang, Z.D.:
Implementing topological quantum manipulation with superconducting circuits.
Phys. Rev. A \textbf{79}, 040303 (2009)


\bibitem{xue3}   Xue, Z.-Y.:
Simulation of anyonic fractional statistics of Kitaev¡¯s toric model
in circuit QED.  EPL \textbf{93},  20007 (2011)

\bibitem{s3} Zhou, L., Gong, Z.R.,  Liu, Y.-x., Sun, C.P.,  Nori, F.:
Controllable scattering of photons inside a one-dimensional resonator waveguide.
Phys. Rev. Lett. \textbf{101}, 100501 (2008)

\bibitem{s4} Zhou, L., Dong, H.,  Liu, Y.-x., Sun, C.P.,  Nori, F.:
Quantum super-cavity with atomic mirrors. Phys. Rev. A \textbf{78}, 063827 (2008)

\bibitem{s5} Zhou, L., Yang, S.,  Liu, Y.-x., Sun, C.P.,  Nori, F.:
Quantum Zeno switch for single-photon coherent transport. Phys. Rev. A \textbf{80}, 062109 (2009)

\bibitem{s6} Liao, J.-Q.,  Gong, Z.R., Zhou, L., Liu, Y.-x., Sun, C.P.,  Nori, F.:
Controlling the transport of single photons by tuning the frequency of either one or two cavities in an array of coupled cavities.
Phys. Rev. A \textbf{81}, 042304 (2010)

\bibitem{you}  You, J.Q.,  Nori, F.:
Quantum information processing with superconducting qubits in a microwave field, Phys. Rev. B \textbf{68}, 064509 (2003)

\bibitem{zhu1}  Zhu, S.-L., Wang, Z.D.,  Yang, K.:
Quantum-information processing using Josephson junctions coupled through cavities. Phys. Rev. A \textbf{68},
034303 (2003)

\bibitem{zhu2}  Zhu, S.-L., Wang, Z.D., Zanardi, P.: Geometric quantum computation and multiqubit entanglement
with superconducting qubits inside a cavity. Phys. Rev. Lett. \textbf{94}, 100502 (2005)

\bibitem{xue4} Xue, Z.-Y.,Wang, Z.D.: Simple unconventional geometric scenario of one-way quantum computation
with superconducting qubits inside a cavity. Phys. Rev. A \textbf{75}, 064303 (2007)

\bibitem{A.Wallraff} Wallraff, A., Schuster, D.I.,  Blais, A., Frunzio,  L.,  Huang, R.S., Majer, J., Kumar, S., Girvin, S.M., Schoelkopf, R.:
Strong coupling of a single photon to a superconducting qubit using circuit quantum electrodynamics.
Nature  (London) \textbf{431}, 162 (2004)
%


\bibitem{ion2} Zhu, S.-L.,  Monroe, C.,  Duan,  L.-M.:
Trapped ion quantum computation with transverse phonon modes.
Phys. Rev. Lett. \textbf{97}, 050505 (2006).

\bibitem{A.Blais} Blais, A., Huang, R.S., Wallraff, A., Girvin, S.M., Schoelkopf, R.J.:
Cavity quantum electrodynamics for superconducting electrical circuits: an architecture for quantum computation.
Phys. Rev. A  \textbf{69}, 062320 (2004).
%


\bibitem{xue} Xue, Z.-Y.: Fast geometric gate operation of superconducting charge qubits in circuit QED. Quantum
Inf. Process. \textbf{11}, 1381-1388 (2012)

\bibitem{s} Solano, E., Agarwal, G.S.,  Walther, H.:
Strong-driving-assisted multipartite entanglement in cavity QED.
Phys. Rev. Lett. \textbf{90}, 027903 (2003)

\bibitem{aw2}  Leek, P.J.,  Baur, M., Fink, J.M., Bianchetti, R., Steffen, L., Filipp, S.,  Wallraff, A.:
Cavity quantum electrodynamics with separate photon storage and qubit readout modes.
Phys. Rev. Lett. \textbf{104}, 100504 (2010)

\bibitem{D.F.} Walls, D.F., Milburn,  G.J.: Quantum Optics. Springer, Berlin (2008)

\bibitem{Mohan} Sarovar, M., Goan, H.-S., Spiller, T.P.,  Milburn, G.J.:
High-fidelity measurement and quantum feedback control in circuit QED.
Phys. Rev. A \textbf{72}, 062327 (2005)



\end{thebibliography}
\end{document}